\documentclass[%
 reprint,showpacs,showkeys,
 amsmath,amssymb,
 aps,
]{revtex4-1}

\usepackage{graphicx}
\usepackage{epstopdf}
\usepackage{dcolumn}
\usepackage{bm}
\usepackage{hyperref} 

\begin{document}

\preprint{APS/123-QED}

\title{Lyapunov spectra of Coulombic and gravitational periodic systems}
\author{Pankaj Kumar}
\author{Bruce N. Miller}%
 \email{b.miller@tcu.edu}

\affiliation{
 Department of Physics and Astronomy,\\
 Texas Christian University, Fort Worth, Texas 76129, USA
}

\date{\today}

\begin{abstract}
We compute Lyapunov spectra for Coulombic and gravitational versions of the one-dimensional systems of parallel sheets with periodic boundary conditions. Exact time evolution of tangent-space vectors are derived and are utilized toward computing Lypaunov characteristic exponents using an event-driven algorithm. The results indicate that the energy dependence of the largest Lyapunov exponent emulates that of Kolmogorov-entropy density for each system at different degrees of freedom. Our approach forms an effective and approximation-free tool toward studying the dynamical properties exhibited by the Coulombic and gravitational systems and finds applications in investigating indications of thermodynamic transitions in large versions of the spatially periodic systems.
\end{abstract}

\maketitle

\section{\label{sec:level1}Introduction\protect\\}

One dimensional systems are of great interest to physicists in terms of their intrinsic properties and as a starting point in the analysis of their more-complicated higher-dimensional counterparts (see \cite{Rybicki1971,Miller1997,Miller1998,Lauritzen2013,Kumar2014} and references therein). In the analysis of large systems considered in plasma and gravitational physics, periodic boundary conditions are preferred \cite{Springiel2006,Bertschinger1998,hockney1988,Hernquist1991} and have been utilized in the study of one-dimensional plasma and gravitational systems \cite{Kunz1974,Schotte1980,miller2010,Kumar2014}. Such studies often rely on numerical simulations to validate the predictions made by the theory, and in cases where theoretical relations have not been mathematically formulated, numerical simulations serve as a powerful approach in characterizing the dynamical behaviors and thermodynamic properties \cite{Kumar2014}.

Numerical studies usually employ the molecular-dynamics (MD) approach in the study of the dynamical systems that undergo phase-space mixing to exhibit ergodic-like behavior. Phase-space mixing is a necessary condition for equilibrium statistical mechanics to apply and is often characterized by the existence of positive Lyapunov characteristic exponents (LCEs) \cite{Krylov1979}. LCEs represent the average rates of exponential divergence of nearby trajectories from a reference trajectory in difference directions of the phase space and quantify the degree of chaos in a dynamical system \cite{Pesin1976,Pesin1977,Benettin1979,ott2002,sprott2003}. In addition, LCEs have also been reported to serve as indicators of phase transitions \cite{Butera1987,Caiani1997,Casetti2000,Dellago1996,Barre2001}.

Numerical calculation of LCEs may be realized by studying the geometry of the phase-space trajectories \cite{Caiani1997,Casetti2000} for smooth systems. In general, however, if the time evolution of each particles' position and velocity can be followed for a system, the largest LCE may be calculated by finding the rate of divergence between a reference trajectory and a nearby test trajectory obtained by pertubing the former \cite{ott2002}. This numerical approach was extended to the case of systems with periodic boundary conditions by Kumar and Miller and was applied to find the largest LCE for a spatially-periodic one-dimensional Coulombic system \cite{Kumar2014}.

While the largest LCE is a good indicator of the degree of maximum chaotic instability in a system, the mixing speed, that is, the rate at which a given phase-space volume element of the phase space diffuses across the allowed regions of the phase space is indicated by the Kolmogorov-Sinai (KS) entropy.  For ergodic-like Hamiltonian systems, KS entropy is obtained as the sum of the positive LCEs \cite{Pesin1976}. In simulation, full spectrum of LCEs may be obtained by finding the time-averaged exponential rates of growth of perturbation vectors applied to the phase-space flows in the tangent space \cite{Benettin1979,Posch1997,Posch1998,Tsuchiya2000}. An exact numerical method of calculating the full Lyapunov spectrum was proposed for the case of one-dimensional gravitation gas \cite{Benettin1979}.

Even though a full spectrum is highly desirable, the calculation becomes computationally challenging for systems with large degrees of freedom and it is usually impractical to aim for a full spectrum through $N$-body simulations \cite{Benettin1979}. In this paper, we extend the numerical approach presented in Ref. \cite{Benettin1979} to compute the complete Lyapunov spectra  of spatially-periodic one-dimensional Coulombic and gravitational systems and show that the energy dependence of the largest LCE emulates that of the sum of all the positive LCEs for the two versions of the system.

The paper is organized as follows: In section \ref{sec:Model}, we describe the Coulombic and gravitational versions of the model and discuss the potential interactions as well as their implications on the phase-space characteristics. In Sec. \ref{sec:LCE}, we recall some theoretical results for LCEs and the general numerical approach for their numerical computation. Section \ref{sec:NBS} presents derivations for the time evolution of the tangent-vectors and the results of the $N$-body simulations. Finally, in Sec. \ref{sec:LCE}, we discuss the results and provide concluding remarks.

\section{Model} \label{sec:Model}
We consider two versions of a spatially periodic lineal system on an $x$-axis with the primitive cell extending in $[-L,L)$ and which contains $N$ infinite sheets, each with a surface mass density $m$. The positions of the sheets are given by $x_1,\hdots,x_N$ with respect to the center of the cell and their corresponding velocities by $v_1,\hdots,v_N$. In one version, the sheets are uncharged and are only interacting gravitationally. The other version is essentially a one-dimensional Coulombic system in which the sheets are charged with a surface charge density $q$ and are immersed in a uniformly distributed negative background such that the net charge is zero. For the case of the charged version of the system, we neglect the gravitational effects and take into account only the Coulomb interactions. If we denote momenta as $p_i \equiv mv_i$, the Hamiltonian of the system may be expressed as
\begin{equation} \label{eq_potential_energy}
\mathcal{H} = \frac{1}{2m}\sum_{i=1}^{N}p^2_i + \kappa \sum_{i<j}^{N} \left(\frac{(x_j-x_i)^2}{2L} - |x_j-x_i|\right),
\end{equation}
where $\kappa = {2\pi kq^2}$ for the case of the Coulombic system \cite{Kumar2014} (with each sheet, henceforth referred to as a \textit{particle} or a \textit{body}, having a surface charge density $q$ in addition to the surface mass density $m$) and $\kappa = -2\pi Gm^2$ for the gravitational system \cite{miller2010}.

\section{Lyapunov Characteristic Exponents} \label{sec:LCE}
\subsection{Theoretical Overview}
Here we provide a brief overview of dynamical system theory that will be helpful in developing the formulations in subsequent sections. While more general and comprehensive discussions are provided in Refs. \cite{Benettin1979, Benettin1980,Sandri1996}, we restrict this overview to a smooth Hamiltonian system. We will see later how the concept may be extended to flows that take place on non-differentiable manifolds.\\

Let the phase-space flow, $\phi^t(z)$ be a one-parameter group of measure-preserving diffeomorphisms $M \to M$, where $M$ is an $n$-dimensional compact differentiable manifold and $z\in M$. For a Hamiltonian system with a phase-space dimensionality of $2N$, $n=2N-1$. If $T_zM$ is the tangent space to $M$ at $z$, then we can define $\mathcal{D}\phi^t_z(w)$ as a linearized flow in the tangent space $(T_zM \to T_{\phi^t_z}M)$, where $w$ is a vector in the tangent space. For a non-zero $w$, there are $n$ independent eigenvectors $e_1,\hdots,e_n$ with $\chi_1,\hdots,\chi_n$ as the corresponding eigenvalues such that $\left|\chi_1\right| \geq \left|\chi_2\right| \geq \cdots \geq \left|\chi_n\right|$. For a periodic orbit with period $t_0$, if we define $\lambda_i \equiv t^{-1}_0\mbox{ln}\left|\chi_i\right|$, then
\begin{equation}
\frac{\left\lVert \mathcal{D}\phi^{kt_0}_z(e_i) \right\lVert}{\left\lVert e_i \right\lVert} = e^{\lambda_ikt_0} \hspace{2pt},
\end{equation}
where $\left\lVert \hspace{6pt} \right\lVert$ represents the Euclidean norm on $T_zM$ and $k$ is a positive integer \cite{Benettin1979}. For a tangent-space vector $w$ with a non-zero component along $e_1$, the divergence for large $t$ will be dominated by $e^{\lambda_1 t}$, and therefore,
\begin{equation} \label{eq_ApproachToLambda1}
\lim_{t \to \infty} \frac{1}{t}\mbox{ln}\frac{\left\lVert \mathcal{D}\phi^{t}_z(w) \right\lVert}{\left\lVert w \right\lVert} = \lambda_1 .
\end{equation} 
$\lambda_1$ is usually called the largest Lyapunov characteristic exponent (LCE) of the orbit represented by the flow $\phi^t(z)$, and is a measure of the overall stability of the orbit; if $\lambda_1 \geq 0$, then the nearby trajectories diverge exponentially. Note that even though we have used a periodic orbit to define the largest LCE, it may be shown that the limit in the left hand side of Eq. (\ref{eq_ApproachToLambda1}) exists and is finite for any given dynamical system and the result applies rather generally under very weak smoothness conditions \cite{Oseledec1968}.\\

LCE defined in Eq. ({\ref{eq_ApproachToLambda1}) may be thought of as the mean exponential growth rate of a one-dimensional ``volume'' (length of a vector $w$) in the tangent space. Therefore, $\lambda_1$ is often referred to as LCE of order $1$. Similarly, $\lambda_p$, that is, LCE of order $p$ (where $1 \leq p \leq n$, $p \in \mathbb{Z}_+$), may be related to the mean exponential rate of growth of a $p$-dimensional hyperparallelepiped formed by the evolution of $p$ linearly independent tangent-space vectors $w_1,\hdots,w_p$. We first find the rate of volume divergence as
\begin{equation}
\lambda^{\mathcal{P}} = \lim_{t \to \infty} \frac{1}{t}\mbox{ln}\frac{{\mbox{Vol}}^{\mathcal{P}} \left[\mathcal{D}\phi^{t}_z(w_1),\hdots,\mathcal{D}\phi^{t}_z(w_p)\right]}{{\mbox{Vol}}^{\mathcal{P}} \left[w_1,\hdots,w_p \right]} ,
\end{equation}
where ${\mbox{Vol}}^{\mathcal{P}}$ represents the volume spanned by a set of $p$ tangent-space vectors. Finally, following Ref. \cite{Oseledec1968}, $\lambda_p$ is found as
\begin{equation} \label{eq_Lambda_p}
\lambda_p = \left\lbrace
\begin{array}{c l}
\lambda^{\mathcal{P}}, & \hspace{5pt} p = 1, \\
\;\\
\lambda^{\mathcal{P}} - \lambda^{\mathcal{P}-1}, & \hspace{5pt} 1 < p \leq n .
\end{array}\right.\;
\end{equation}
\subsection{Numerical Approach}
We start with a randomly chosen set of $n$ orthonormal tangent vectors $\lbrace \hat{w}^0_1,\hdots,\hat{w}^0_n \rbrace$. Clearly, for each $p \leq n$, ${\mbox{Vol}}^{\mathcal{P}}\left[\hat{w}^0_1,\hdots,\hat{w}^0_p \right] = 1$. After a fixed time interval $\tau$, the evolved tangent vectors---which we denote by $\lbrace {w}^1_1,\hdots,{w}^1_n \rbrace$, where ${w}^1_i = \mathcal{D}\phi^{t = \tau}_z(\hat{w}^0_i)$---are, in general, no longer mutually orthogonal. This is because the component of each $\hat{w}^0_i$ along the direction of maximum divergence $e_1$ (that is, $\hat{w}^0_i \cdot e_1 $) will witness a disproportionately larger growth in its value as compared to the remaining components. In order to avoid numerical errors arising from one component getting increasingly large in comparison to the others, a new orthonormal set of tangent vectors $\lbrace \hat{w}^1_1,\hdots,\hat{w}^1_n \rbrace$ is defined after time $\tau$ through Gram-Schmidt reduction on the set of evolved $\lbrace {w}^0_1,\hdots,{w}^0_n \rbrace$. This new set of ornothormal tangent vectors are then used for the following iteration, and the process is recursively repeated until $\lambda^{\mathcal{P}}$ has converged \cite{Benettin1979,Sandri1996}. \\

Numerical calculation of $\lambda^{\mathcal{P}}$ involves finding the corresponding $p$-volume for each iteration. If at the end of the $j$-th iteration, ${w}^j_i= \mathcal{D}\phi^{t = j\tau}_z(\hat{w}^{j-1}_i)$ represent the evolved versions of the orthornormal tangent vectors $\hat{w}^{j-1}_i$, the $p$-volume may be found as the norm of the the exterior product involving the corresponding $p$ vectors, that is,
\begin{equation} \label{eq_VolP_norm}
{\mbox{Vol}}^{\mathcal{P}} [{w}^j_1,\hdots,{w}^j_p] = \left\lVert {w}^j_1\wedge{w}^j_2\wedge\cdots\wedge{w}^j_p \right\lVert .
\end{equation}
Finally, the average exponential growth rate of the $p$-volume is found as
\begin{equation}
\lambda^{\mathcal{P}} = \lim_{l \to \infty} \frac{1}{l\tau}\sum_{j=1}^{l}\left(\mbox{ln}{\mbox{Vol}}^{\mathcal{P}} [{w}^j_1,\hdots,{w}^j_p]\right) ,
\end{equation}
where $l$ is the total number of iterations. A complete set of LCEs $\lbrace \lambda_1,\hdots,\lambda_n\rbrace$, also known as Lyapunov spectrum, may then be obtained for the trajectory by utilizing Eq. (\ref{eq_Lambda_p}) for all permissible values of $p$. Finally, an upper limit on the KS entropy $h_{KS}$ for Hamiltonian systems may be obtained as the sum of positive LCE's \cite{Pesin1976,Benettin1979},\\
\begin{equation} \label{eq_Kentropy}
h_{KS} \leq \lambda_S = \sum_{p=1}^{p_{max}^+} \lambda_p ,
\end{equation}
where ${p_{max}^+}$ is the largest value of $p$ for which $\lambda_p$ is positive and where the equality $h_{KS} = \lambda_S$ holds for ergodic-like systems. The sum of the positive LCEs $\lambda_S$ is often termed as the density of KS entropy.

In the following section, we discuss how we employ the theory and the numerical approach presented thus far to obtain Lyapunov spectra for the Coulombic and gravitational systems discussed in Sec. \ref{sec:Model}. While most of the presented theory applies to the two systems in its original form, non-smoothness arising from the absolute valued linear terms in the potential demand additional consideration. It turns out that, as we shall see, following the time evolution of the tangent-space vectors involves treating the motion as a flow in between two consecutive events of interparticle crossings and as a mapping at each event of such crossings. Consequently, it becomes indispensable to have the ability to find the exact time corresponding to each crossing.

\section{$N$-body Simulation} \label{sec:NBS}

\subsection{Equations of motion}
Positions and velocities of the particles are obtained using event-driven algorithms based on the approaches proposed in Ref. \cite{Kumar2014} for the Coulombic system and in Ref. \cite{miller2010} for the gravitational system. The algorithms employ analytic expressions for the time dependencies of the relative separations $Z_j(t)$ and relative velocities $W_j(t)$ between two consecutive particles in the primitive cell, where $Z_j = (x_{j+1} - x_j)$ and $W_j = (v_{j+1} - v_j)$, with $x_j$ and $v_j$ representing, respectively, the position and velocity of the $j$-th particle whereas $x_{j+1}$ and $v_{j+1}$ representing those of the $(j+1)$-th particle. Combining the results of Refs. \cite{Kumar2014,miller2010}, we find that
\begin{equation} \label{eq_relMotionGrav}
\frac{d}{dt} W_j(t) =  -\frac{\kappa}{m} \left\lbrace \frac{N}{L}Z_j(t) -2\right\rbrace .
\end{equation}
Crossing times may be found by solving $Z_j(t) = 0$ for $t$. The corresponding positions $x_j$ and velocities $v_j$ are obtained using a matrix-inversion subroutine as described in Ref. \cite{Kumar2014}.

\subsection{Time evolution of tangent-space vectors}
In order to follow the time evolution of the tangent vectors,  we adopt an approach based on the ``exact'' numerical method proposed in Ref. \cite{Benettin1979}. The method invokes that, for a one-dimensional Hamiltonian system with $N$ particles, one does not have to restrict to the $(2N-1)$-dimensional manifold $\Gamma_E$. One may alternatively choose to represent the flow $\phi^t$ in the entire $2N$-dimensional phase space (say, $\Omega$) whereby the tangent space $T_z\Gamma_E$ becomes a subspace of $T_z\Omega$.\\  

Let ${z}({x},{v})$ be a point in the phase space $\Omega$, where $x = (x_1,\hdots,x_N)$ and $v = (v_1,\hdots,v_N)$. The equations of motion representing the system are given by
\begin{equation}
{\dot{x}}_j = v_j,
\end{equation}
and
\begin{equation}
{\dot{v}}_j = -\frac{1}{m} \frac{\partial}{\partial x_j}V(x),
\end{equation}
with
\begin{equation} \label{eq_V_x}
V(x) = \kappa \sum_{i<j}^{N} \left(\frac{(x_j-x_i)^2}{2L} - |x_j-x_i|\right).
\end{equation}
Similarly, if we have a vector $w(\xi,\eta)$ in the tangent space $T_z\Omega$, then the variational equations \cite{Benettin1979} governing the evolution of $w$ are given by
\begin{equation} \label{eq_eta_zeta_matrix}
\left(
\begin{matrix}
\dot{\xi}\\
\dot{\eta}
\end{matrix} \right) = \left(
\begin{matrix}
0 & I_N\\
A(x) & 0
\end{matrix} \right) \left(
\begin{matrix}
{\xi}\\
{\eta}
\end{matrix} \right),
\end{equation}
where $I_N$ is the $N\times N$ identity matrix and $A(x)$ is an $N\times N$ matrix whose elements are given by
\begin{equation}
A_{ij}(x) = -\frac{1}{m} \frac{\partial^2}{\partial x_i \partial x_j}V(x).
\end{equation}
For the potential expressed in Eq. (\ref{eq_V_x}), one finds that
\begin{equation} \label{eq_Aij_x}
A_{ij}(x) = \left\lbrace
\begin{array}{l r}
-\frac{\kappa}{m} \left. \left[ -\frac{1}{L} - 2\delta(x_i-x_j) \right] \right., & i \neq j, \\
\;\\
-\frac{\kappa}{m} \left. \left[ \frac{N-1}{L} - \sum\limits_{i\neq k=1}^N 2\delta(x_i-x_k) \right] \right., & i = j. 
\end{array}\right.\;
\end{equation} \label{eq_zeta_j_dot}
Using Eqs. (\ref{eq_eta_zeta_matrix}) and (\ref{eq_Aij_x}), we may deduce that
\begin{equation}
{\dot{\xi}}_j = {\eta}_j,
\end{equation} 
and
\begin{equation} \label{eq_eta_j_dot}
{\dot{\eta}}_j = -\frac{\kappa}{m} \left[\frac{N {\xi}_j - {\Xi}_{S}}{L} - 2\sum\limits_{i\neq j=1}^N (\xi_j - \xi_i) \delta(x_j-x_i) \right],
\end{equation}
where
\begin{equation}
\Xi_S = \sum\limits_{j=1}^N \xi_j \hspace{2pt}.
\end{equation}
Equations (\ref{eq_zeta_j_dot}) and (\ref{eq_eta_j_dot}) imply that
\begin{equation} \label{eqTime_Evol_xi}
\frac{d^2 \xi_j}{dt^2} = -\frac{\kappa}{m} \left[\frac{N {\xi}_j - {\Xi}_{S}}{L} - 2\sum\limits_{i\neq j=1}^N (\xi_j - \xi_i) \delta(x_j-x_i) \right].
\end{equation}
Time evolution of $\Xi_S$ between the events of interparticle crossings may be deduced by adding Eq. \ref{eqTime_Evol_xi} for all values of $j$ as
\begin{equation}
\sum\limits_{j=1}^N \frac{d^2 \xi_j}{dt^2} = 0,
\end{equation}
which implies that
\begin{equation} \label{eq_Xi_diff_eq}
\frac{d^2 \Xi_S}{dt^2} = 0 .
\end{equation}
Solution to Eq. (\ref{eq_Xi_diff_eq}) yields
\begin{equation}
\Xi_S(t) = H_S(0)t + \Xi_S(0) .
\end{equation}
where
\begin{equation}
H_S = \sum\limits_{j=1}^N \eta_j .
\end{equation}
Hence, in between the events of crossings, the evolution of tangent vectors takes the form
\begin{equation} \label{eqTime_Evol_xi_2}
\frac{d^2 \xi_j}{dt^2} = -\frac{\kappa}{m} \left[\frac{N {\xi}_j - H_S(0)t + \Xi_S(0)}{L} \right].
\end{equation}
With the two values of $\kappa$, one for the Coulombic system and the other for the gravitational system, one may solve Eq. (\ref{eqTime_Evol_xi_2}) to find the exact dependencies of $\xi_j$ and $\eta_j$ on time.
\subsubsection{Coulombic system}
Utilizing the initial conditions for the Coulombic system with $\kappa = 2\pi kq^2$, we obtain the solutions to Eq. \ref{eqTime_Evol_xi_2} for $\xi_j$ and $\eta_j$ in between crossings as
\begin{eqnarray}
\xi_j(t) = \frac{H_S(0)}{N}t\hspace{4pt} +&& \hspace{4pt} \frac{\Xi_S(0)}{N} + \frac{1}{\omega}\left\lbrace \eta_j(0) - \frac{H_S(0)}{N} \right\rbrace \sin{\omega t}\nonumber \\
&& + \hspace{4pt} \left\lbrace {\xi_j(0)} - \frac{\Xi_S(0)}{N} \right\rbrace\cos{\omega t},
\end{eqnarray}
and,
\begin{eqnarray}
\eta_j(t) = \frac{H_S(0)}{N} + \left\lbrace \eta_j(0) - \frac{H_S(0)}{N} \right\rbrace \cos{\omega t} \nonumber \\
- \hspace{4pt} \omega \left\lbrace {\xi_j(0)} - \frac{\Xi_S(0)}{N} \right\rbrace \sin{\omega t} ,
\end{eqnarray}
where $\omega \equiv \sqrt{\frac{\kappa N}{mL}} = \sqrt{\frac{2\pi kq^2N}{mL}}$.

If the $r$-th and $s$-th particles undergo a crossing at time $t=t_c$, and  $t^-$ and $t^+$ respectively denote the instants just before and after $t=t_c$, then
\begin{equation}
\eta_r(t^+) = \eta_r(t^-) + \frac{4\pi kq^2}{m}\frac{\left(\xi_r(t^-)-\xi_s(t^-)\right)}{|v_r(t^-)-v_s(t^-)|},
\end{equation}
\begin{equation}
\eta_s(t^+) = \eta_s(t^-) - \frac{4\pi kq^2}{m}\frac{\left(\xi_r(t^-)-\xi_s(t^-)\right)}{|v_r(t^-)-v_s(t^-)|}.
\end{equation}

\subsubsection{Gravitational system}
For the gravitational system, $\kappa = -2\pi Gm^2$. Similar to the Coulombic case, we utilize the initial conditions to find solutions to Eq. \ref{eqTime_Evol_xi_2} for $\xi_i$ and $\eta_j$ as functions of time for the gravitational case in between the events of crossings: 
\begin{equation}
\xi_j(t) = \frac{1}{2\Lambda}\lbrace A_g e^{\Lambda t} + B_g e^{-\Lambda t} \rbrace + \frac{H_S(0)}{N}t + \frac{\Xi_S(0)}{N},
\end{equation}
\begin{equation}
\eta_j(t) = \frac{1}{2}\lbrace A_g e^{\Lambda t} - B_g e^{-\Lambda t} \rbrace + \frac{H_S(0)}{N},
\end{equation}
where,
\begin{equation}
A_g = \Lambda \xi_j(0) + \eta_j(0) - \frac{H_S(0)}{N} - \Lambda \frac{\Xi_S(0)}{N},
\end{equation}
\begin{equation}
B_g = \Lambda \xi_j(0) - \eta_j(0) + \frac{H_S(0)}{N} - \Lambda \frac{\Xi_S(0)}{N},
\end{equation}
and $\Lambda \equiv \sqrt{\frac{kN}{mL}} = \sqrt{\frac{2\pi GmN}{L}}$.
For the $r$-th and $s$-th particle involved in a crossing, we get
\begin{equation}
\eta_r(t^+) = \eta_r(t^-) - {4\pi Gm}\frac{\left(\xi_r(t^-)-\xi_s(t^-)\right)}{|v_r(t^-)-v_s(t^-)|},
\end{equation}
\begin{equation}
\eta_s(t^+) = \eta_s(t^-) + {4\pi Gm}\frac{\left(\xi_r(t^-)-\xi_s(t^-)\right)}{|v_r(t^-)-v_s(t^-)|},
\end{equation}
where we have used the same definitions of $t_c$, $t^-$, and $t^+$ as we did in the expressions' Coulombic counterparts.

\subsection{$p$-volume and Lyapunov spectrum}
We perform numerical computations using algorithms that are driven by tracking the events of interparticle crossings. Since the time derivatives of the velocities of the particles involved in a crossing undergo abrupt changes, tracking of crossing becomes indispensable.  In other words, even if we were to sample the positions and velocities at fixed intervals of time, we would still need to track every single event of crossing \cite{Benettin1979}. Therefore, instead of choosing fixed time intervals for Gram-Schmidt orthonormalization of the tangent-space vectors, we choose to do it after each crossing. It should be noted that the duration of each iteration, fixed or variable, is irrelevant as long it remains short enough for the various components of the tangent vectors to remain comparable with a given precision offered by the computing platform utilized.\\

After an iteration, say, the $(j-1)$-th iteration, ending at time $t = t^{j-1}_c$, each of the $2N$ orthonormal tangent vectors $\hat{w}^{j-1}_1,\hdots,\hat{w}^{j-1}_N$ is allowed to evolve for the duration $\delta t^{j}_c$ of the $j$-th iteration. $\delta t^{j}_c$ may be thought of as the time elapsed between the instants right after the $(j-1)$-th crossing and right before the $j$-th crossing.  Then the evolved vectors may be given by ${w}^{j}_i= \mathcal{D}\phi^{t^j_c}_z(\hat{w}^{j-1}_i)$, where
\begin{equation} \label{eq_t_j_c}
t^j_c = t^{j-1}_c + \delta t^{j}_c = \delta t^{1}_c + \delta t^{2}_c +\cdots+ \delta t^j_c.
\end{equation}

For the $j$-th iteration, $p$-volume is found as follows: We define a $p \times p$ symmetric matrix $\mathcal{G}^j_p$ whose elements ${\mathcal{G}^j_{p}}_{\mu \nu}$ are inner products between ${w}^{j}_\mu$ and ${w}^{j}_\nu$, where $1 \leq \mu,\nu \leq p$ (with $\mu,\nu \in \mathbb{Z}_+$). That is,
\begin{equation}
\mathcal{G}^j_p = \left(
\begin{matrix}
\langle {w}^{j}_1,{w}^{j}_1 \rangle & \langle {w}^{j}_1,{w}^{j}_2  \rangle & \cdots & \langle {w}^{j}_1,{w}^{j}_p \rangle\\
\langle {w}^{j}_2,{w}^{j}_1 \rangle & \langle {w}^{j}_2,{w}^{j}_2  \rangle & \cdots & \langle {w}^{j}_2,{w}^{j}_p \rangle\\
\vdots           & \vdots           & \ddots & \vdots\\
\langle {w}^{j}_p,{w}^{j}_1 \rangle & \langle {w}^{j}_p,{w}^{j}_2  \rangle & \cdots & \langle {w}^{j}_p,{w}^{j}_p \rangle   
\end{matrix} \right).
\end{equation}
The matrix $\mathcal{G}^j_p$, also referred to as \textit{Gram matrix}, encapsulates the necessary geometric information about the subspace spanned by the set of vectors $\lbrace {w}^{j}_1,\hdots,{w}^{j}_p \rbrace$ such as the lengths of the vectors and the angles between them. The absolute value of the determinant of $\mathcal{G}^j_p$, known as the \textit{Gramian}, is essentially the square of the norm of the exterior product \cite{Shilov2012}. Using Eq. (\ref{eq_VolP_norm}), the $p$-volume is found as
\begin{equation}
{\mbox{Vol}}^{\mathcal{P}} [{w}^j_1,\hdots,{w}^j_p] = \sqrt{\left| \mbox{det}\left(\mathcal{G}^j_p\right)\right|}.
\end{equation}
With the ability to find each $p$-volume for a given iteration, the final value of the corresponding $\lambda^{\mathcal{P}}$ are found as
\begin{equation} \label{eq_lambdapl}
\lambda^{\mathcal{P}} = \lim_{l \to \infty} \frac{1}{t^l_c}\sum_{j=1}^{l}\mbox{ln}\left({\mbox{Vol}}^{\mathcal{P}} [{w}^j_1,\hdots,{w}^j_p]\right) ,
\end{equation}
where $t^l_c$, as defined in Eq. (\ref{eq_t_j_c}), is the total time elapsed for $l$ crossings to occur. Finally, LCEs $\lambda_p$ and Kolmogorov-entropy density $\lambda_S$ are obtained using Eqs. (\ref{eq_Lambda_p}) and (\ref{eq_Kentropy}) respectively.

\begin{figure*}[t]
\begin{center} 
    \resizebox{0.49\textwidth}{!}{\includegraphics{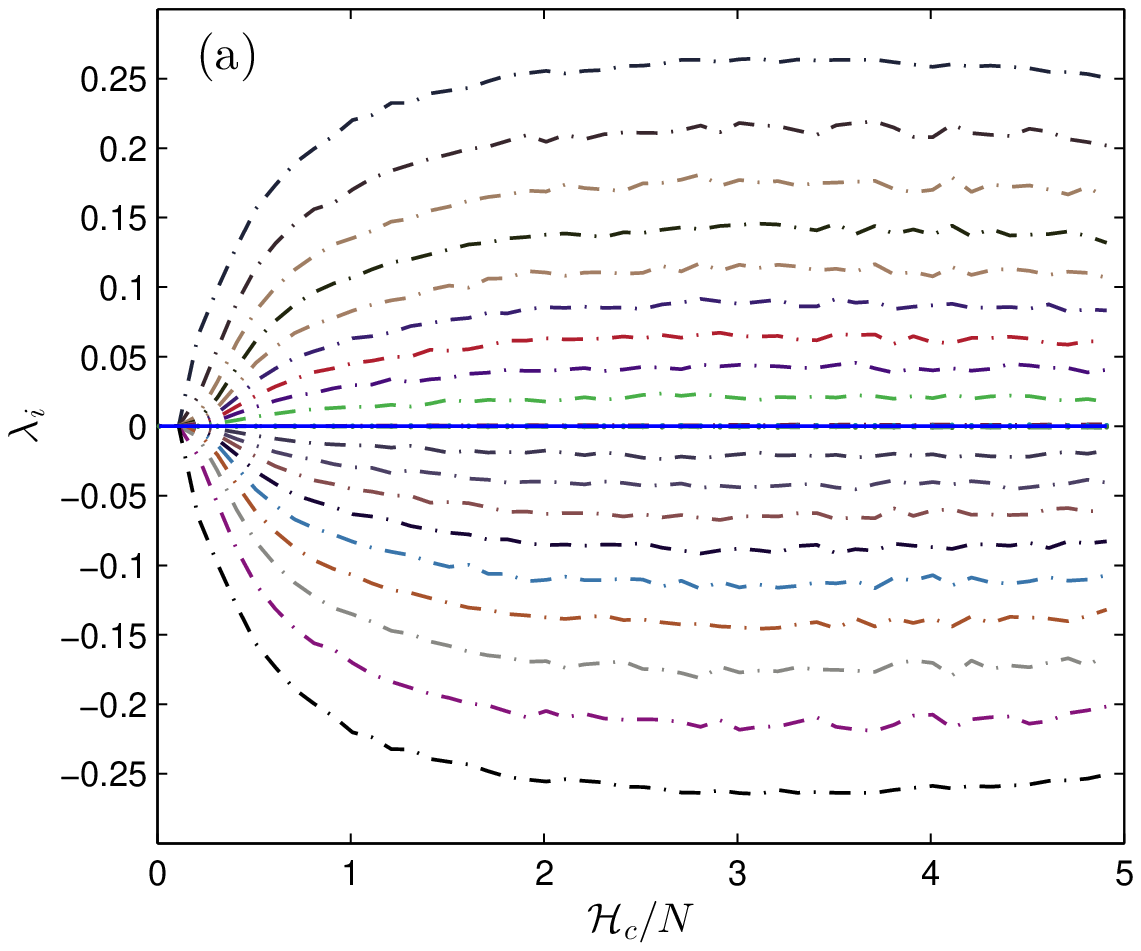}} 
    \resizebox{0.49\textwidth}{!}{\includegraphics{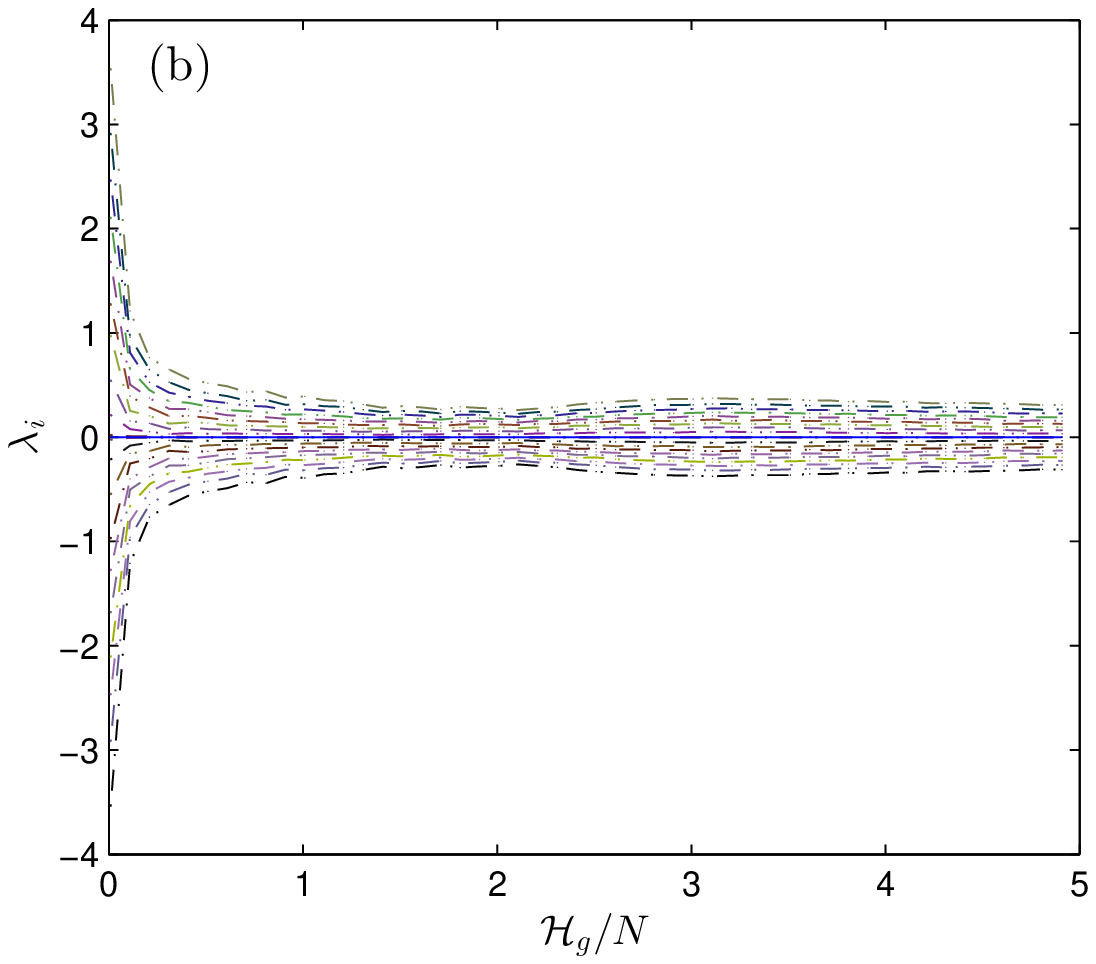}}
\end{center}
\caption{\label{fig:Spectrum_11_C_G} Full spectra of LCEs plotted against per-particle energy for (a) Coulombic system, and (b) gravitational system, with $N=11$. The topmost curve shows $\lambda_{1}$, the second to top curve shows $\lambda_2$, and so on all the way to the curve on the very bottom representing $\lambda_{22}$ in Figs. (a) and (b). The central solid (blue) line in each plot shows the sum of LCEs. $\mathcal{H}_c$ and $\mathcal{H}_g$ are expressed in units of $\frac{2L}{N}|\kappa|$. $\lambda_i$ are expressed in units of (a) $\omega$, and (b) $\Lambda$.} 
\end{figure*}

\begin{figure*}[t]
\begin{center} 
    \resizebox{0.49\textwidth}{!}{\includegraphics{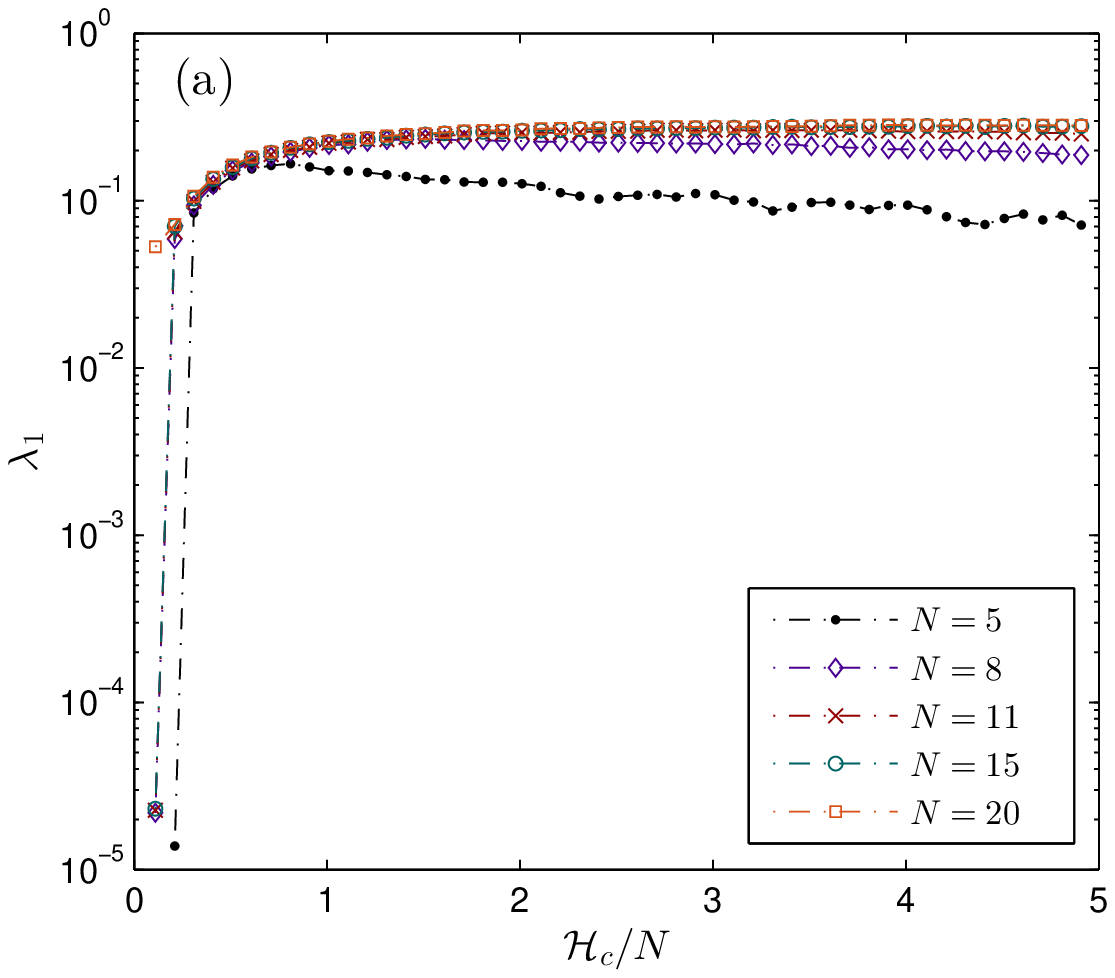}} 
    \resizebox{0.49\textwidth}{!}{\includegraphics{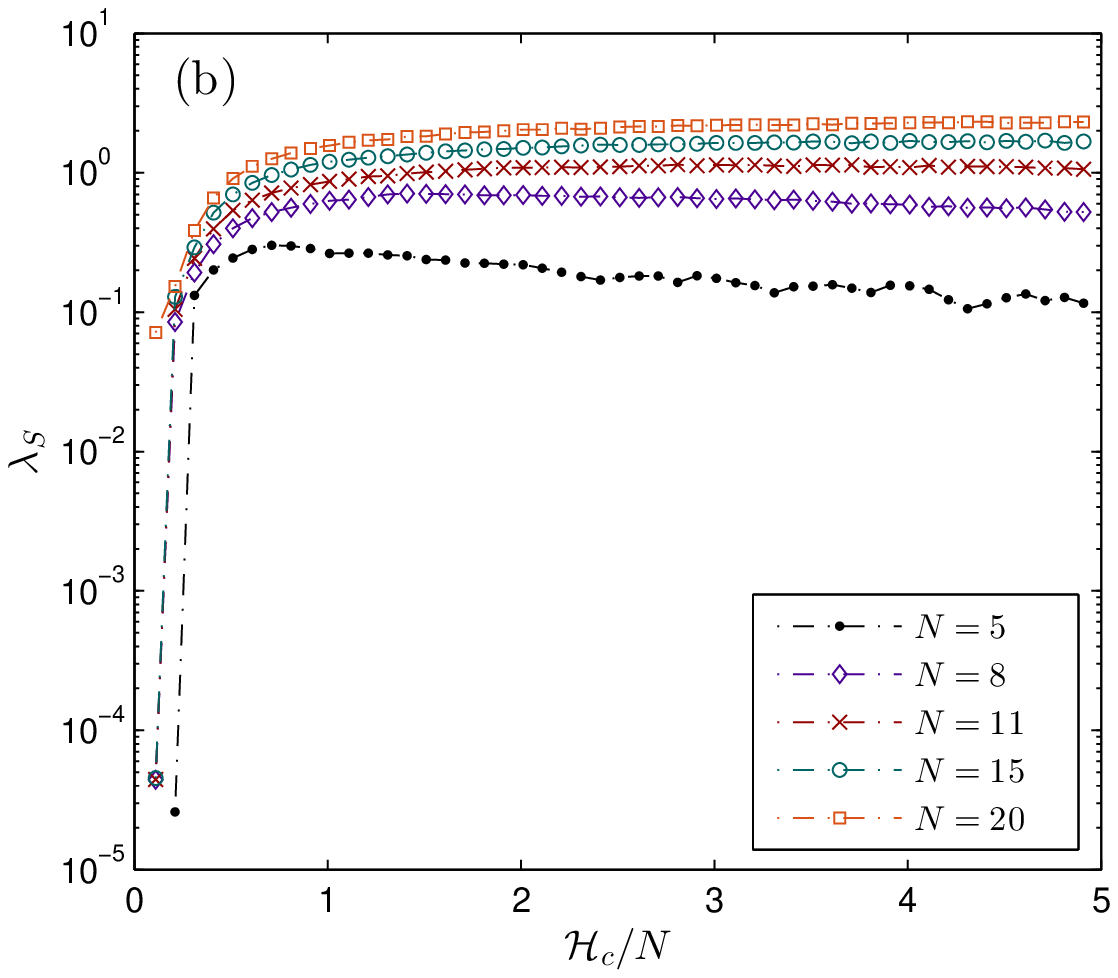}}
\end{center}
\caption{\label{fig:Coul_LambdaN} Energy dependence of (a) the largest LCE, and (b) Kolmogorov-entropy density for Coulombic system with different degrees of freedom. $\lambda_1$ and $\lambda_S$ are expressed in units of $\omega$ whereas $\mathcal{H}_c$ in units of $\frac{2L}{N}|\kappa|$.}
\end{figure*}

\begin{figure*}[t]
\begin{center} 
    \resizebox{0.49\textwidth}{!}{\includegraphics{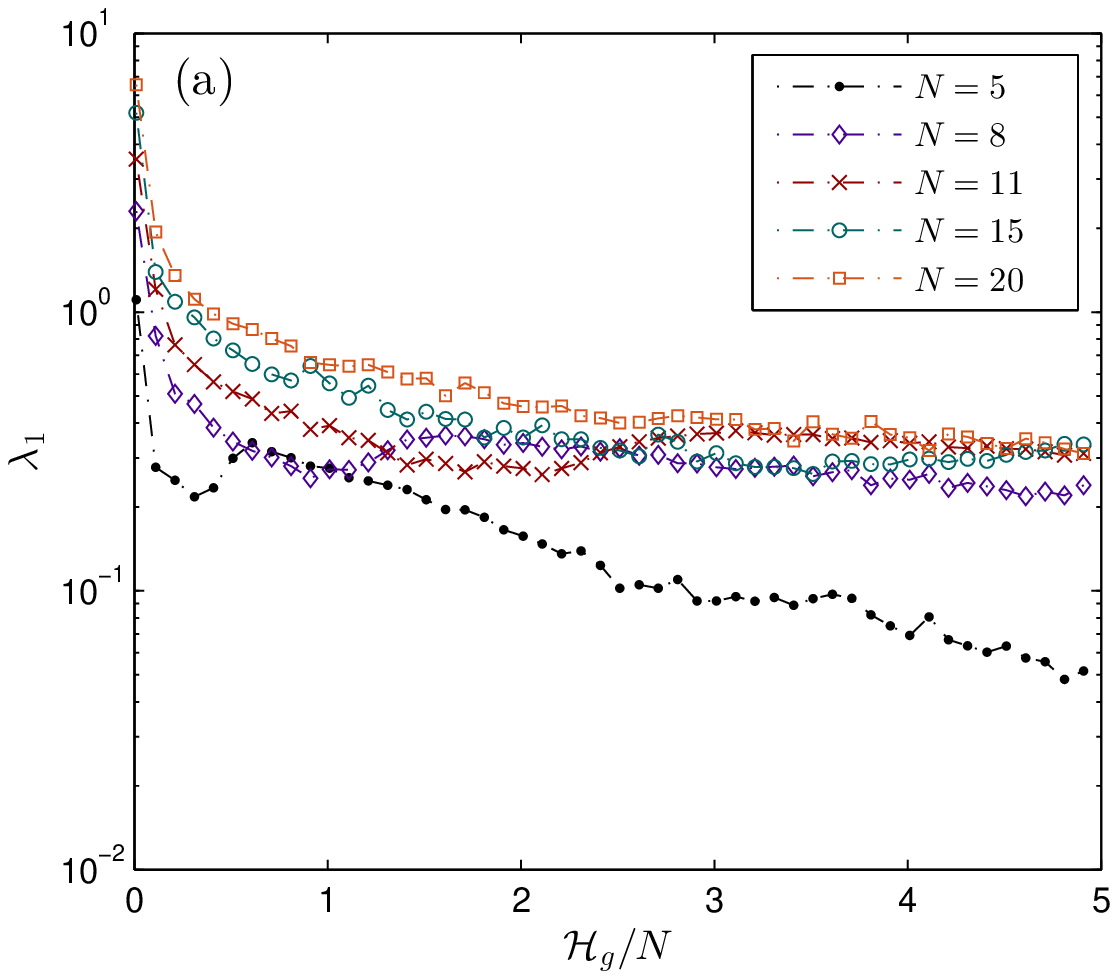}} 
    \resizebox{0.49\textwidth}{!}{\includegraphics{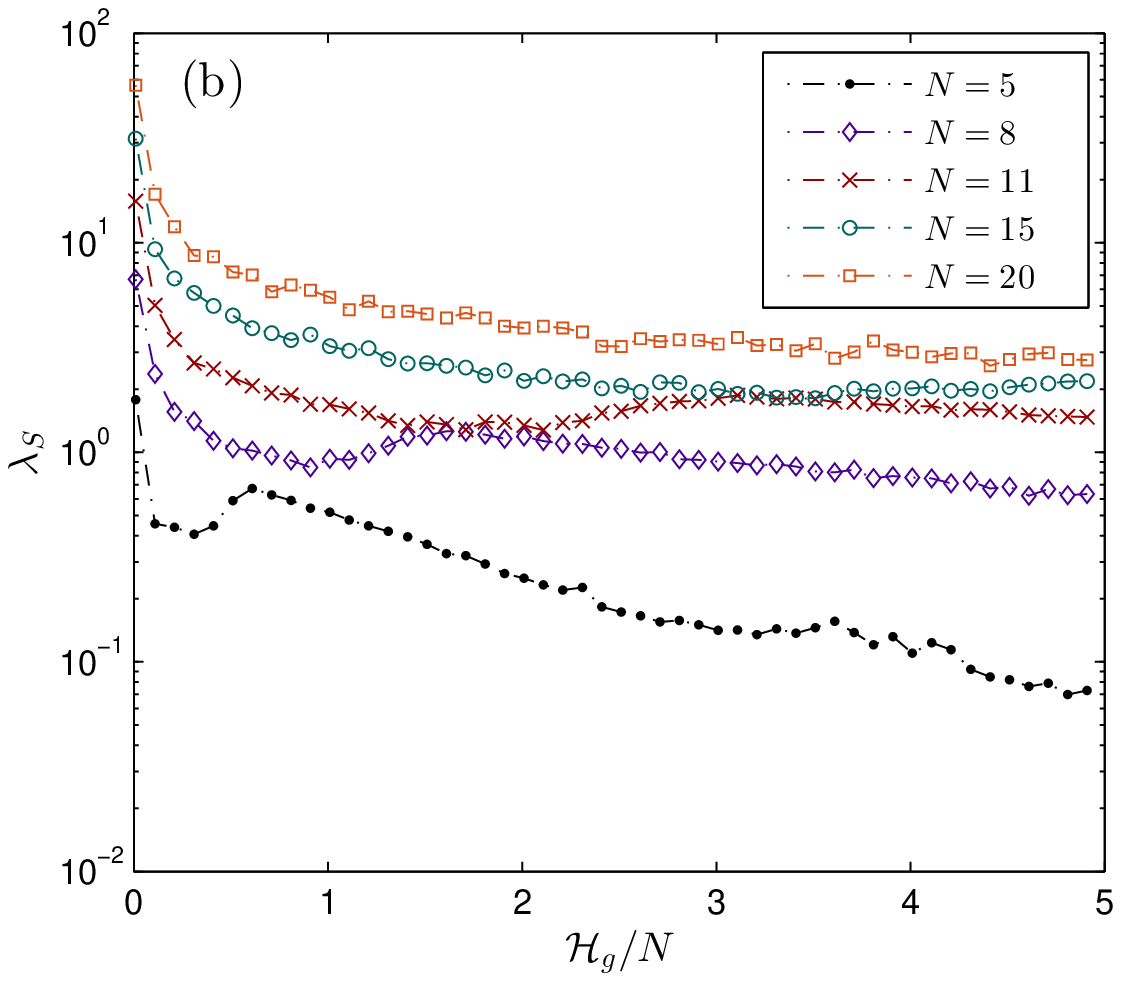}}
\end{center}
\caption{\label{fig:Grav_LambdaN} Energy dependence of (a) the largest LCE, and (b) Kolmogorov-entropy density for gravitational system with different degrees of freedom. $\lambda_1$ and $\lambda_S$ are expressed in units of $\Lambda$ whereas $\mathcal{H}_g$ in units of $\frac{2L}{N}|\kappa|$..} 
\end{figure*}

\begin{figure*}[t]
\begin{center} 
    \resizebox{0.49\textwidth}{!}{\includegraphics{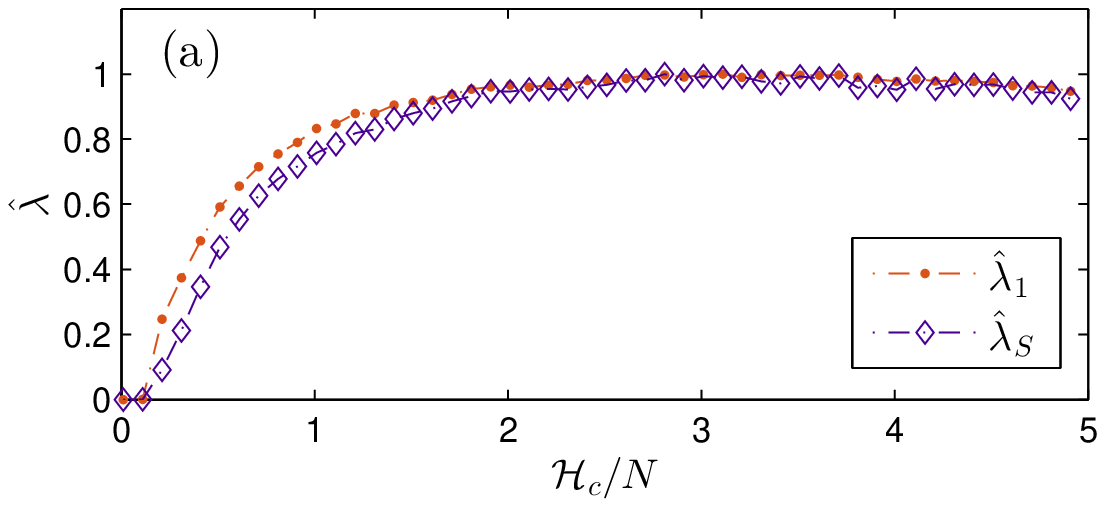}} 
    \resizebox{0.49\textwidth}{!}{\includegraphics{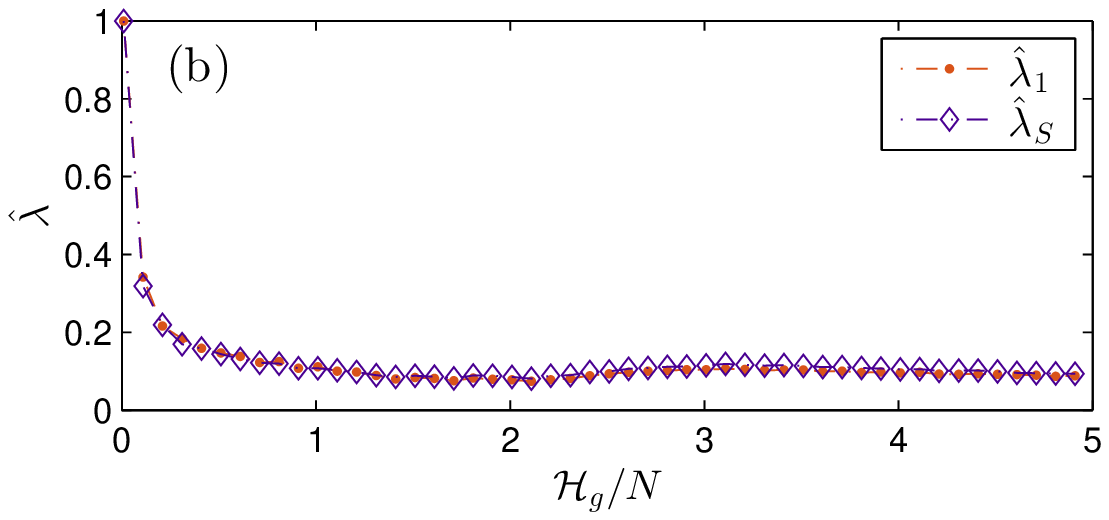}}
\end{center}
\caption{\label{fig:Largest_Sum_11_C_G} Energy dependence of the normalized values of the largest LCE and Kolmogorov-entropy density for (a) Coulombic system, and (b) gravitational system for $N=11$. $\mathcal{H}_c$ and $\mathcal{H}_g$ are expressed in units of $\frac{2L}{N}|\kappa|$ whereas $\hat{\lambda}$ are dimensionless. } 
\end{figure*}

\subsection{Results}
In our simulation, the initial conditions are chosen as follows: For a given number of particles $N$ and per-particle energy $\mathcal{H}$, if $\mathcal{H}$ is lower than the maximum allowed value of the potential energy $V_{max}$, then the particle positions are chosen randomly such that the potential energy is slightly smaller than the target value of $\mathcal{H}$. For $\mathcal{H}$ greater than $V_{max}$, the particle positions are randomly selected such that the potential energy is close to $V_{max}$.  Velocities are chosen randomly from a Gaussian distribution and are scaled such that the sum of the potential and kinetic energies exactly equals the target value of $\mathcal{H}$ \cite{Kumar2014}.

Simulations are performed by rescaling the system parameters in a system of dimensionless units such that the number density $N/2L = 1$ and the characteristic frequencies $\omega$, $\Lambda$ equal unity \citep{miller2010,Kumar2014}. Energies $\mathcal{H}_c$ and $\mathcal{H}_g$ (respectively for Coulombic and gravitational systems) are measured with respect to the minimum values of the corresponding potential energies allowed for each $N$.

Before running the Lyapunov algorithm, we allowed the system to evolve for a relaxation period of $500$ time units (in terms of $1/\omega$ or $1/\Lambda$). In the calculation of LCEs, the total number of iterations $l$ for each $\lambda^{\mathcal{P}}$ in Eq. \ref{eq_lambdapl} are decided by an adaptive algorithm which runs a minimum preliminary number of iterations assigned beforehand and then continues running until the value of $\lambda^{\mathcal{P}}$ has converged to within a pre-specified tolerance for the standard deviation. The tolerance value that we specified was $0.1$ percent of the mean from the newest $500,000$ iterations, with a minimum of $1$ million iterations. In each run, we found that the values converged to within our specified tolerance after the preliminary run of $1$ million iterations.

We computed Lyapunov spectra for the Coulombic and gravitational systems with varying number of particles $N$ with $5 \leq N \leq 20$. Figure \ref{fig:Spectrum_11_C_G} shows examples of the dependence of the various LCEs on the per-particle energy for the Coulombic and gravitational systems with $N=11$. The figure also shows the sum of all LCEs for each case. As expected, the sums were found to be close to zero.

The energy dependencies of the largest LCE and the Kolmogorov-entropy density with $N=5,8,11,15,$ and $20$ have been shown for the Coulombic and gravitational systems in Figs. \ref{fig:Coul_LambdaN} and \ref{fig:Grav_LambdaN} respectively. It can be seen in each case that the behavior of $\lambda_S$ versus $\mathcal{H}/N$ resembles, in general, a scaled version of $\lambda_1$ versus $\mathcal{H}/N$. To elucidate this, we have presented comparative plots of the normalized versions, ${\hat{\lambda}}_1$ and ${\hat{\lambda}}_S$, of $\lambda_1$ and $\lambda_S$ against per-particle energy for $N=11$ in Fig. \ref{fig:Largest_Sum_11_C_G}, where we have divided $\lambda_1$ and $\lambda_S$ by their respective maximum values to get the normalized values.

\section{Discussion and Conclusions} \label{sec:DandC}
Our study provides interesting insight into the chaotic dynamics of the two versions of the periodic system under the conditions of varying energy and degrees of freedom. The results of the Coulombic system are consistent those provided in Ref. \cite{Kumar2014}. $\lambda_1$ for the Coulombic system stays zero as long as the energy is low enough for the particles to not undergo crossings. As the energy is increased from a low value, $\lambda_1$ sees an initial increase, reaches a maximum and then decreases as the energy is progressively raised. However, with an increase in the number of particles, the right edge of ``hill'' near the maximum opens up to form a ``plateau'' and flattens out asymptotically as shown in Fig. \ref{fig:Coul_LambdaN}(a). Interestingly, our results also suggest that all the other LCEs are, more or less, scaled (and inverted, for the case of negative LCEs) versions of the largest LCE, as we can see in Fig. \ref{fig:Spectrum_11_C_G}(a) for $N=11$.

Unlike the Coulombic version, the gravitational system shows a maximum degree of chaos at low energies. Our results suggest that the largest LCE starts off with a high value at low energies and decreases to a minimum as the energy is increased for a given $N$. With further rise in energy, $\lambda_1$ increases, reaches and maximum and then decreases asymptotically. As the number of particles is increased, the local trough near the local minimum and the hill near the local maximum start opening up to the right with the asymptotic edge rising up to a form a plateau as shown in Fig. \ref{fig:Grav_LambdaN}(b). Moreover, it can also be seen in Fig. \ref{fig:Grav_LambdaN}(b) that the local minimum and maximum themselves shift to the right as well with an increase in the number of particles. It should be noted that in the thermodynamic limit, that is, for versions of the system with sufficiently large $N$, the LCEs may exhibit discontinuities in their values or their slopes near the troughs and crests when plotted against temperature. Such an observation may indicate toward existence of phase transitions \cite{Dellago1996,Barre2001}.

While the energy dependencies of $\lambda_1$ are drastically different for the two versions of the system, the other LCEs for the gravitational system are also, in general, scaled versions of $\lambda_1$, similar to the Coulombic system, as exemplified in Fig. \ref{fig:Spectrum_11_C_G}(b) for $N=11$. Moreover, in both versions of the spatially-periodic system, the energy dependence of LCEs tends to approach a common limiting behavior as $N$ is increased. This is in contrast to the free-boundary gravitating case in which the values of LCEs were shown to increase linearly with an increase in the number of particles \cite{Benettin1979}.

From a dynamical perspective, the results are consistent with the theoretical predictions for Hamiltonian systems \cite{Benettin1979} as outlined below:

\textbf{(a)}. The sum of LCEs was found to converge to zero for all energies and $N$.

\textbf{(b)}. For the ordered set $\lbrace\lambda_i\rbrace$ ($\lambda_1\geq\lambda_2\geq\hdots\geq\lambda_{2N}$), the results show that
\begin{equation}
\lambda_i \sim -\lambda_{2N-i+1}, \hspace{15pt} i=1,2,\hdots,2N .
\end{equation}

\textbf{(c)}. In addition, our results show that
\begin{equation}
\lambda_{N-1} \sim \lambda_N \sim \lambda_{N+1} \sim \lambda_{N+2} \sim 0.
\end{equation}

Result \textbf{(c)} is nothing but a consequence of the conservation of momentum \cite{Benettin1979}.

To summarize, we have provided an exact method to compute full spectra of LCEs for the spatially periodic versions of the one-dimensional Coulombic and gravitational systems. Analytic expressions for time-evolutions of the tangent vectors were derived and used in numerically computing LCEs using an efficient event-driven algorithm. While the resulting values of the largest LCE for the Coulombic system agree with reported previously \cite{Kumar2014} , our exact approach offers striking advantages over the method used in Ref.\cite{Kumar2014} in that it allows one to calculate a full spectrum of LCEs rather than just the largest LCE. Second, the results of the exact method do not depend on the size of the perturbation. In finding the largest LCE using finite perturbations to a reference orbit as discussed in Ref. \cite{Kumar2014}, one has to first make sure that the value chosen for initial perturbation is small enough. This becomes challenging for systems in which particles tend stay clumped together. An example of such behavior is seen in the gravitational system at low energies. Finally, the exact approach circumvents the need for defining a test trajectory altogether, which as discussed in Ref. \cite{Kumar2014} poses difficulties in expressing phase-space separations for systems with periodic boundary conditions. Nevertheless, the method discussed in Ref. \cite{Kumar2014} still remains powerful, and perhaps the only resort, in dealing with spatially-periodic systems for which analytic evolution of tangent-space vectors may not be obtained.

The results of our study also indicate that the energy dependence of the largest LCE captures the general behavior of the dependence of the Kolmogorov-entropy density on energy for both Coulombic and gravitational systems. This result is particularly significant because of the numerical difficulties encountered  while calculating the full Lyapunov spectra of large systems. For the two versions of the spatially-periodic system, our study suggests that one may gain insights into a full spectrum of LCEs by simply looking at the largest LCE, thereby allowing one to evade the computational complications faced when calculating a full spectrum. 

It should be noted that, for a given number of particles, the energy dependence of the LCEs roughly followed the same behavior for any randomly selected initial conditions with only slight deviations. As the number of particles was increased, the deviations became smaller leading the behaviors to converge to a single universal one, indicating toward the approach of ergodic-like nature. Moreover, the convergence times of the LCNs for different randomly-chosen initial conditions also showed uniformity for larger number of particles, thereby pointing toward a consistent relaxation to equilibrium with increasing degrees of freedom. However, the exact dependence of relaxation time on the number of particles requires further investigation and we plan to pursue it in our future work.

Finally, it is worth emphasizing that if a phase transition occurs in either of two spatially-periodic systems, the \textit{temperature dependence} of the largest LCE is expected to show a transitioning behavior in the thermodynamic limit (large $N$-limit) \cite{Bonasera1995,Dellago1996,Barre2001}. In our future work, we plan to utilize the various previosuly reported numerical techniques \cite{Kumar2014} as well as the approach presented in this paper to examine the chaotic and thermodynamic properties of the periodic gravitational and Coulombic systems for indications of phase transitions.



\begin{acknowledgments}
The authors thank Dr. Harald Posch of University of Vienna and Dr. Igor Prokhorenkov of Texas Christian University for valuable insights and helpful discussions.
\end{acknowledgments}

\bibliography{KumarMiller_ChaoticDynamicsVer1}

\providecommand{\noopsort}[1]{}\providecommand{\singleletter}[1]{#1}%
\begin{thebibliography}{31}%
\makeatletter
\providecommand \@ifxundefined [1]{%
 \@ifx{#1\undefined}
}%
\providecommand \@ifnum [1]{%
 \ifnum #1\expandafter \@firstoftwo
 \else \expandafter \@secondoftwo
 \fi
}%
\providecommand \@ifx [1]{%
 \ifx #1\expandafter \@firstoftwo
 \else \expandafter \@secondoftwo
 \fi
}%
\providecommand \natexlab [1]{#1}%
\providecommand \enquote  [1]{``#1''}%
\providecommand \bibnamefont  [1]{#1}%
\providecommand \bibfnamefont [1]{#1}%
\providecommand \citenamefont [1]{#1}%
\providecommand \href@noop [0]{\@secondoftwo}%
\providecommand \href [0]{\begingroup \@sanitize@url \@href}%
\providecommand \@href[1]{\@@startlink{#1}\@@href}%
\providecommand \@@href[1]{\endgroup#1\@@endlink}%
\providecommand \@sanitize@url [0]{\catcode `\\12\catcode `\$12\catcode
  `\&12\catcode `\#12\catcode `\^12\catcode `\_12\catcode `\%12\relax}%
\providecommand \@@startlink[1]{}%
\providecommand \@@endlink[0]{}%
\providecommand \url  [0]{\begingroup\@sanitize@url \@url }%
\providecommand \@url [1]{\endgroup\@href {#1}{\urlprefix }}%
\providecommand \urlprefix  [0]{URL }%
\providecommand \Eprint [0]{\href }%
\providecommand \doibase [0]{http://dx.doi.org/}%
\providecommand \selectlanguage [0]{\@gobble}%
\providecommand \bibinfo  [0]{\@secondoftwo}%
\providecommand \bibfield  [0]{\@secondoftwo}%
\providecommand \translation [1]{[#1]}%
\providecommand \BibitemOpen [0]{}%
\providecommand \bibitemStop [0]{}%
\providecommand \bibitemNoStop [0]{.\EOS\space}%
\providecommand \EOS [0]{\spacefactor3000\relax}%
\providecommand \BibitemShut  [1]{\csname bibitem#1\endcsname}%
\let\auto@bib@innerbib\@empty
\bibitem [{\citenamefont {Rybicki}(1971)}]{Rybicki1971}%
  \BibitemOpen
  \bibfield  {author} {\bibinfo {author} {\bibfnamefont {G.~B.}\ \bibnamefont
  {Rybicki}},\ }\href {\doibase 10.1007/BF00649195} {\bibfield  {journal}
  {\bibinfo  {journal} {Astrophysics and Space Science}\ }\textbf {\bibinfo
  {volume} {14}},\ \bibinfo {pages} {56} (\bibinfo {year} {1971})}\BibitemShut
  {NoStop}%
\bibitem [{\citenamefont {Yawn}\ and\ \citenamefont
  {Miller}(1997)}]{Miller1997}%
  \BibitemOpen
  \bibfield  {author} {\bibinfo {author} {\bibfnamefont {K.~R.}\ \bibnamefont
  {Yawn}}\ and\ \bibinfo {author} {\bibfnamefont {B.~N.}\ \bibnamefont
  {Miller}},\ }\href {\doibase 10.1103/PhysRevLett.79.3561} {\bibfield
  {journal} {\bibinfo  {journal} {Phys. Rev. Lett.}\ }\textbf {\bibinfo
  {volume} {79}},\ \bibinfo {pages} {3561} (\bibinfo {year}
  {1997})}\BibitemShut {NoStop}%
\bibitem [{\citenamefont {Miller}\ and\ \citenamefont
  {Youngkins}(1998)}]{Miller1998}%
  \BibitemOpen
  \bibfield  {author} {\bibinfo {author} {\bibfnamefont {B.~N.}\ \bibnamefont
  {Miller}}\ and\ \bibinfo {author} {\bibfnamefont {P.}~\bibnamefont
  {Youngkins}},\ }\href {\doibase 10.1103/PhysRevLett.81.4794} {\bibfield
  {journal} {\bibinfo  {journal} {Phys. Rev. Lett.}\ }\textbf {\bibinfo
  {volume} {81}},\ \bibinfo {pages} {4794} (\bibinfo {year}
  {1998})}\BibitemShut {NoStop}%
\bibitem [{\citenamefont {Lauritzen}\ \emph {et~al.}(2013)\citenamefont
  {Lauritzen}, \citenamefont {Gustainis},\ and\ \citenamefont
  {Mann}}]{Lauritzen2013}%
  \BibitemOpen
  \bibfield  {author} {\bibinfo {author} {\bibfnamefont {A.}~\bibnamefont
  {Lauritzen}}, \bibinfo {author} {\bibfnamefont {P.}~\bibnamefont
  {Gustainis}}, \ and\ \bibinfo {author} {\bibfnamefont {R.~B.}\ \bibnamefont
  {Mann}},\ }\href
  {http://scitation.aip.org/content/aip/journal/jmp/54/7/10.1063/1.4815834}
  {\bibfield  {journal} {\bibinfo  {journal} {Journal of Mathematical Physics}\
  }\textbf {\bibinfo {volume} {54}},\ \bibinfo {eid} {072703} (\bibinfo {year}
  {2013})}\BibitemShut {NoStop}%
\bibitem [{\citenamefont {Kumar}\ and\ \citenamefont
  {Miller}(2014)}]{Kumar2014}%
  \BibitemOpen
  \bibfield  {author} {\bibinfo {author} {\bibfnamefont {P.}~\bibnamefont
  {Kumar}}\ and\ \bibinfo {author} {\bibfnamefont {B.~N.}\ \bibnamefont
  {Miller}},\ }\href {\doibase 10.1103/PhysRevE.90.062918} {\bibfield
  {journal} {\bibinfo  {journal} {Phys. Rev. E}\ }\textbf {\bibinfo {volume}
  {90}},\ \bibinfo {pages} {062918} (\bibinfo {year} {2014})}\BibitemShut
  {NoStop}%
\bibitem [{\citenamefont {Springiel}\ \emph {et~al.}(2006)\citenamefont
  {Springiel}, \citenamefont {Frenk},\ and\ \citenamefont
  {White}}]{Springiel2006}%
  \BibitemOpen
  \bibfield  {author} {\bibinfo {author} {\bibfnamefont {V.}~\bibnamefont
  {Springiel}}, \bibinfo {author} {\bibfnamefont {C.~S.}\ \bibnamefont
  {Frenk}}, \ and\ \bibinfo {author} {\bibfnamefont {S.~D.~M.}\ \bibnamefont
  {White}},\ }\href@noop {} {\bibfield  {journal} {\bibinfo  {journal}
  {Nature}\ }\textbf {\bibinfo {volume} {440}},\ \bibinfo {pages} {1137}
  (\bibinfo {year} {2006})}\BibitemShut {NoStop}%
\bibitem [{\citenamefont {Bertschinger}(1998)}]{Bertschinger1998}%
  \BibitemOpen
  \bibfield  {author} {\bibinfo {author} {\bibfnamefont {E.}~\bibnamefont
  {Bertschinger}},\ }\href@noop {} {\bibfield  {journal} {\bibinfo  {journal}
  {Annual Review of Astronomy and Astrophysics}\ }\textbf {\bibinfo {volume}
  {36}},\ \bibinfo {pages} {599} (\bibinfo {year} {1998})}\BibitemShut
  {NoStop}%
\bibitem [{\citenamefont {Hockney}\ and\ \citenamefont
  {Eastwood}(1988)}]{hockney1988}%
  \BibitemOpen
  \bibfield  {author} {\bibinfo {author} {\bibfnamefont {R.~W.}\ \bibnamefont
  {Hockney}}\ and\ \bibinfo {author} {\bibfnamefont {J.~W.}\ \bibnamefont
  {Eastwood}},\ }\href {http://books.google.com/books?id=nTOFkmnCQuIC} {\emph
  {\bibinfo {title} {Computer Simulation Using Particles}}}\ (\bibinfo
  {publisher} {Taylor \& Francis},\ \bibinfo {year} {1988})\BibitemShut
  {NoStop}%
\bibitem [{\citenamefont {Hernquist}\ \emph {et~al.}(1991)\citenamefont
  {Hernquist}, \citenamefont {Bouchet},\ and\ \citenamefont
  {Suto}}]{Hernquist1991}%
  \BibitemOpen
  \bibfield  {author} {\bibinfo {author} {\bibfnamefont {L.}~\bibnamefont
  {Hernquist}}, \bibinfo {author} {\bibfnamefont {F.~R.}\ \bibnamefont
  {Bouchet}}, \ and\ \bibinfo {author} {\bibfnamefont {Y.}~\bibnamefont
  {Suto}},\ }\href {\doibase 10.1086/191530} {\bibfield  {journal} {\bibinfo
  {journal} {Astrophysical Journal Supplement}\ }\textbf {\bibinfo {volume}
  {75}},\ \bibinfo {pages} {231} (\bibinfo {year} {1991})}\BibitemShut
  {NoStop}%
\bibitem [{\citenamefont {Kunz}(1974)}]{Kunz1974}%
  \BibitemOpen
  \bibfield  {author} {\bibinfo {author} {\bibfnamefont {H.}~\bibnamefont
  {Kunz}},\ }\href {\doibase http://dx.doi.org/10.1016/0003-4916(74)90413-8}
  {\bibfield  {journal} {\bibinfo  {journal} {Annals of Physics}\ }\textbf
  {\bibinfo {volume} {85}},\ \bibinfo {pages} {303 } (\bibinfo {year}
  {1974})}\BibitemShut {NoStop}%
\bibitem [{\citenamefont {Schotte}\ and\ \citenamefont
  {Truong}(1980)}]{Schotte1980}%
  \BibitemOpen
  \bibfield  {author} {\bibinfo {author} {\bibfnamefont {K.~D.}\ \bibnamefont
  {Schotte}}\ and\ \bibinfo {author} {\bibfnamefont {T.~T.}\ \bibnamefont
  {Truong}},\ }\href {\doibase 10.1103/PhysRevA.22.2183} {\bibfield  {journal}
  {\bibinfo  {journal} {Phys. Rev. A}\ }\textbf {\bibinfo {volume} {22}},\
  \bibinfo {pages} {2183} (\bibinfo {year} {1980})}\BibitemShut {NoStop}%
\bibitem [{\citenamefont {Miller}\ and\ \citenamefont
  {Rouet}(2010)}]{miller2010}%
  \BibitemOpen
  \bibfield  {author} {\bibinfo {author} {\bibfnamefont {B.~N.}\ \bibnamefont
  {Miller}}\ and\ \bibinfo {author} {\bibfnamefont {J.-L.}\ \bibnamefont
  {Rouet}},\ }\href@noop {} {\bibfield  {journal} {\bibinfo  {journal}
  {Physical Review E}\ }\textbf {\bibinfo {volume} {82}},\ \bibinfo {pages}
  {066203} (\bibinfo {year} {2010})}\BibitemShut {NoStop}%
\bibitem [{\citenamefont {Krylov}\ \emph {et~al.}(1979)\citenamefont {Krylov},
  \citenamefont {Migdal}, \citenamefont {Sinai},\ and\ \citenamefont
  {Zeeman}}]{Krylov1979}%
  \BibitemOpen
  \bibfield  {author} {\bibinfo {author} {\bibfnamefont {N.}~\bibnamefont
  {Krylov}}, \bibinfo {author} {\bibfnamefont {A.}~\bibnamefont {Migdal}},
  \bibinfo {author} {\bibfnamefont {Y.~G.}\ \bibnamefont {Sinai}}, \ and\
  \bibinfo {author} {\bibfnamefont {Y.~L.}\ \bibnamefont {Zeeman}},\
  }\href@noop {} {\bibfield  {journal} {\bibinfo  {journal} {Works on the
  Foundations of Statistical Physics by Nikolai Sergeevich Krylov translsated
  by AB Migdal, Ya. G. Sinai and Yu. L. Zeeman. Princeton Series in Physics.
  Published by Princeton University Press, Princeton, New Jersey 1979.}\ }
  (\bibinfo {year} {1979})}\BibitemShut {NoStop}%
\bibitem [{\citenamefont {PESIN}(1976)}]{Pesin1976}%
  \BibitemOpen
  \bibfield  {author} {\bibinfo {author} {\bibfnamefont {I.}~\bibnamefont
  {PESIN}},\ }in\ \href@noop {} {\emph {\bibinfo {booktitle} {Akademiia Nauk
  SSSR, Doklady}}},\ Vol.\ \bibinfo {volume} {226}\ (\bibinfo {year} {1976})\
  pp.\ \bibinfo {pages} {774--777}\BibitemShut {NoStop}%
\bibitem [{\citenamefont {Pesin}(1977)}]{Pesin1977}%
  \BibitemOpen
  \bibfield  {author} {\bibinfo {author} {\bibfnamefont {Y.~B.}\ \bibnamefont
  {Pesin}},\ }\href@noop {} {\bibfield  {journal} {\bibinfo  {journal} {Russian
  Mathematical Surveys}\ }\textbf {\bibinfo {volume} {32}},\ \bibinfo {pages}
  {55} (\bibinfo {year} {1977})}\BibitemShut {NoStop}%
\bibitem [{\citenamefont {Benettin}\ \emph {et~al.}(1979)\citenamefont
  {Benettin}, \citenamefont {Froeschle},\ and\ \citenamefont
  {Scheidecker}}]{Benettin1979}%
  \BibitemOpen
  \bibfield  {author} {\bibinfo {author} {\bibfnamefont {G.}~\bibnamefont
  {Benettin}}, \bibinfo {author} {\bibfnamefont {C.}~\bibnamefont {Froeschle}},
  \ and\ \bibinfo {author} {\bibfnamefont {J.~P.}\ \bibnamefont
  {Scheidecker}},\ }\href {\doibase 10.1103/PhysRevA.19.2454} {\bibfield
  {journal} {\bibinfo  {journal} {Phys. Rev. A}\ }\textbf {\bibinfo {volume}
  {19}},\ \bibinfo {pages} {2454} (\bibinfo {year} {1979})}\BibitemShut
  {NoStop}%
\bibitem [{\citenamefont {Ott}(2002)}]{ott2002}%
  \BibitemOpen
  \bibfield  {author} {\bibinfo {author} {\bibfnamefont {E.}~\bibnamefont
  {Ott}},\ }\href@noop {} {\emph {\bibinfo {title} {Chaos in Dynamical
  Systems}}}\ (\bibinfo  {publisher} {Cambridge University Press},\ \bibinfo
  {year} {2002})\ pp.\ \bibinfo {pages} {137--145}\BibitemShut {NoStop}%
\bibitem [{\citenamefont {Sprott}(2003)}]{sprott2003}%
  \BibitemOpen
  \bibfield  {author} {\bibinfo {author} {\bibfnamefont {J.}~\bibnamefont
  {Sprott}},\ }\href@noop {} {\emph {\bibinfo {title} {Chaos and Time-series
  Analysis}}}\ (\bibinfo  {publisher} {Oxford University Press},\ \bibinfo
  {year} {2003})\ pp.\ \bibinfo {pages} {116--117}\BibitemShut {NoStop}%
\bibitem [{\citenamefont {Butera}\ and\ \citenamefont
  {Caravati}(1987)}]{Butera1987}%
  \BibitemOpen
  \bibfield  {author} {\bibinfo {author} {\bibfnamefont {P.}~\bibnamefont
  {Butera}}\ and\ \bibinfo {author} {\bibfnamefont {G.}~\bibnamefont
  {Caravati}},\ }\href {\doibase 10.1103/PhysRevA.36.962} {\bibfield  {journal}
  {\bibinfo  {journal} {Phys. Rev. A}\ }\textbf {\bibinfo {volume} {36}},\
  \bibinfo {pages} {962} (\bibinfo {year} {1987})}\BibitemShut {NoStop}%
\bibitem [{\citenamefont {Caiani}\ \emph {et~al.}(1997)\citenamefont {Caiani},
  \citenamefont {Casetti}, \citenamefont {Clementi},\ and\ \citenamefont
  {Pettini}}]{Caiani1997}%
  \BibitemOpen
  \bibfield  {author} {\bibinfo {author} {\bibfnamefont {L.}~\bibnamefont
  {Caiani}}, \bibinfo {author} {\bibfnamefont {L.}~\bibnamefont {Casetti}},
  \bibinfo {author} {\bibfnamefont {C.}~\bibnamefont {Clementi}}, \ and\
  \bibinfo {author} {\bibfnamefont {M.}~\bibnamefont {Pettini}},\ }\href
  {\doibase 10.1103/PhysRevLett.79.4361} {\bibfield  {journal} {\bibinfo
  {journal} {Phys. Rev. Lett.}\ }\textbf {\bibinfo {volume} {79}},\ \bibinfo
  {pages} {4361} (\bibinfo {year} {1997})}\BibitemShut {NoStop}%
\bibitem [{\citenamefont {Casetti}\ \emph {et~al.}(2000)\citenamefont
  {Casetti}, \citenamefont {Pettini},\ and\ \citenamefont
  {Cohen}}]{Casetti2000}%
  \BibitemOpen
  \bibfield  {author} {\bibinfo {author} {\bibfnamefont {L.}~\bibnamefont
  {Casetti}}, \bibinfo {author} {\bibfnamefont {M.}~\bibnamefont {Pettini}}, \
  and\ \bibinfo {author} {\bibfnamefont {E.~G.~D.}\ \bibnamefont {Cohen}},\
  }\href {\doibase http://dx.doi.org/10.1016/S0370-1573(00)00069-7} {\bibfield
  {journal} {\bibinfo  {journal} {Physics Reports}\ }\textbf {\bibinfo {volume}
  {337}},\ \bibinfo {pages} {237} (\bibinfo {year} {2000})}\BibitemShut
  {NoStop}%
\bibitem [{\citenamefont {Dellago}\ and\ \citenamefont
  {Posch}(1996)}]{Dellago1996}%
  \BibitemOpen
  \bibfield  {author} {\bibinfo {author} {\bibfnamefont {C.}~\bibnamefont
  {Dellago}}\ and\ \bibinfo {author} {\bibfnamefont {H.~A.}\ \bibnamefont
  {Posch}},\ }\href {\doibase http://dx.doi.org/10.1016/0378-4371(96)00069-6}
  {\bibfield  {journal} {\bibinfo  {journal} {Physica A: Statistical Mechanics
  and its Applications}\ }\textbf {\bibinfo {volume} {230}},\ \bibinfo {pages}
  {364} (\bibinfo {year} {1996})}\BibitemShut {NoStop}%
\bibitem [{\citenamefont {Barre}\ and\ \citenamefont
  {Dauxois}(2001)}]{Barre2001}%
  \BibitemOpen
  \bibfield  {author} {\bibinfo {author} {\bibfnamefont {J.}~\bibnamefont
  {Barre}}\ and\ \bibinfo {author} {\bibfnamefont {T.}~\bibnamefont
  {Dauxois}},\ }\href {http://stacks.iop.org/0295-5075/55/i=2/a=164} {\bibfield
   {journal} {\bibinfo  {journal} {Europhysics Letters}\ }\textbf {\bibinfo
  {volume} {55}},\ \bibinfo {pages} {164} (\bibinfo {year} {2001})}\BibitemShut
  {NoStop}%
\bibitem [{\citenamefont {Dellago}\ and\ \citenamefont
  {Posch}(1997)}]{Posch1997}%
  \BibitemOpen
  \bibfield  {author} {\bibinfo {author} {\bibfnamefont {C.}~\bibnamefont
  {Dellago}}\ and\ \bibinfo {author} {\bibfnamefont {H.}~\bibnamefont
  {Posch}},\ }\href@noop {} {\bibfield  {journal} {\bibinfo  {journal} {Physica
  A: Statistical Mechanics and its Applications}\ }\textbf {\bibinfo {volume}
  {240}},\ \bibinfo {pages} {68} (\bibinfo {year} {1997})}\BibitemShut
  {NoStop}%
\bibitem [{\citenamefont {Milanovi\ifmmode~\acute{c}\else \'{c}\fi{}}\ \emph
  {et~al.}(1998)\citenamefont {Milanovi\ifmmode~\acute{c}\else \'{c}\fi{}},
  \citenamefont {Posch},\ and\ \citenamefont {Thirring}}]{Posch1998}%
  \BibitemOpen
  \bibfield  {author} {\bibinfo {author} {\bibfnamefont {L.}~\bibnamefont
  {Milanovi\ifmmode~\acute{c}\else \'{c}\fi{}}}, \bibinfo {author}
  {\bibfnamefont {H.~A.}\ \bibnamefont {Posch}}, \ and\ \bibinfo {author}
  {\bibfnamefont {W.}~\bibnamefont {Thirring}},\ }\href {\doibase
  10.1103/PhysRevE.57.2763} {\bibfield  {journal} {\bibinfo  {journal} {Phys.
  Rev. E}\ }\textbf {\bibinfo {volume} {57}},\ \bibinfo {pages} {2763}
  (\bibinfo {year} {1998})}\BibitemShut {NoStop}%
\bibitem [{\citenamefont {Tsuchiya}\ and\ \citenamefont
  {Gouda}(2000)}]{Tsuchiya2000}%
  \BibitemOpen
  \bibfield  {author} {\bibinfo {author} {\bibfnamefont {T.}~\bibnamefont
  {Tsuchiya}}\ and\ \bibinfo {author} {\bibfnamefont {N.}~\bibnamefont
  {Gouda}},\ }\href {\doibase 10.1103/PhysRevE.61.948} {\bibfield  {journal}
  {\bibinfo  {journal} {Phys. Rev. E}\ }\textbf {\bibinfo {volume} {61}},\
  \bibinfo {pages} {948} (\bibinfo {year} {2000})}\BibitemShut {NoStop}%
\bibitem [{\citenamefont {Benettin}\ \emph {et~al.}(1980)\citenamefont
  {Benettin}, \citenamefont {Galgani}, \citenamefont {Giorgilli},\ and\
  \citenamefont {Strelcyn}}]{Benettin1980}%
  \BibitemOpen
  \bibfield  {author} {\bibinfo {author} {\bibfnamefont {G.}~\bibnamefont
  {Benettin}}, \bibinfo {author} {\bibfnamefont {L.}~\bibnamefont {Galgani}},
  \bibinfo {author} {\bibfnamefont {A.}~\bibnamefont {Giorgilli}}, \ and\
  \bibinfo {author} {\bibfnamefont {J.-M.}\ \bibnamefont {Strelcyn}},\ }\href
  {\doibase 10.1007/BF02128236} {\bibfield  {journal} {\bibinfo  {journal}
  {Meccanica}\ }\textbf {\bibinfo {volume} {15}},\ \bibinfo {pages} {9}
  (\bibinfo {year} {1980})}\BibitemShut {NoStop}%
\bibitem [{\citenamefont {Sandri}(1996)}]{Sandri1996}%
  \BibitemOpen
  \bibfield  {author} {\bibinfo {author} {\bibfnamefont {M.}~\bibnamefont
  {Sandri}},\ }\href@noop {} {\bibfield  {journal} {\bibinfo  {journal}
  {Mathematica Journal}\ }\textbf {\bibinfo {volume} {6}},\ \bibinfo {pages}
  {78} (\bibinfo {year} {1996})}\BibitemShut {NoStop}%
\bibitem [{\citenamefont {Oseledec}(1968)}]{Oseledec1968}%
  \BibitemOpen
  \bibfield  {author} {\bibinfo {author} {\bibfnamefont {V.~I.}\ \bibnamefont
  {Oseledec}},\ }\href@noop {} {\bibfield  {journal} {\bibinfo  {journal}
  {Trans. Moscow Math. Soc}\ }\textbf {\bibinfo {volume} {19}},\ \bibinfo
  {pages} {197} (\bibinfo {year} {1968})}\BibitemShut {NoStop}%
\bibitem [{\citenamefont {Shilov}\ and\ \citenamefont
  {Silverman}(2012)}]{Shilov2012}%
  \BibitemOpen
  \bibfield  {author} {\bibinfo {author} {\bibfnamefont {G.~E.}\ \bibnamefont
  {Shilov}}\ and\ \bibinfo {author} {\bibfnamefont {R.~A.}\ \bibnamefont
  {Silverman}},\ }\href@noop {} {\emph {\bibinfo {title} {An introduction to
  the theory of linear spaces}}}\ (\bibinfo  {publisher} {Courier
  Corporation},\ \bibinfo {year} {2012})\BibitemShut {NoStop}%
\bibitem [{\citenamefont {Bonasera}\ \emph {et~al.}(1995)\citenamefont
  {Bonasera}, \citenamefont {Latora},\ and\ \citenamefont
  {Rapisarda}}]{Bonasera1995}%
  \BibitemOpen
  \bibfield  {author} {\bibinfo {author} {\bibfnamefont {A.}~\bibnamefont
  {Bonasera}}, \bibinfo {author} {\bibfnamefont {V.}~\bibnamefont {Latora}}, \
  and\ \bibinfo {author} {\bibfnamefont {A.}~\bibnamefont {Rapisarda}},\ }\href
  {\doibase 10.1103/PhysRevLett.75.3434} {\bibfield  {journal} {\bibinfo
  {journal} {Phys. Rev. Lett.}\ }\textbf {\bibinfo {volume} {75}},\ \bibinfo
  {pages} {3434} (\bibinfo {year} {1995})}\BibitemShut {NoStop}%
\end{thebibliography}%
\end{document}